\documentclass{article}

\usepackage{graphicx}

\newcommand{\be}{\begin{equation}}
\newcommand{\ee}{\end{equation}}
\newcommand{\nn}{\mbox{} \nonumber \\ \mbox{} }
\newcommand{\ba}{\begin{eqnarray}}
\newcommand{\ea}{\end{eqnarray}}
\newcommand{\om}{\omega}
\newcommand{\Alfven}{Alfv\'{e}n }

\newcommand{\B}{{\bf B}}

\newcommand{\J}{{\bf J}}

\renewcommand{\k}{{\bf k}}

\newcommand\eg{\textit{e.g.\ }}

\newcommand{\NS}{neutron star\,}
\newcommand{\NSs}{{neutron stars\,}}
\newcommand{\ms}{{magnetosphere\,}}

\begin{document}

\title{Dissipation of magnetic fields in neutron star crusts due to development of a  tearing mode}
\author{Maxim Lyutikov}

\maketitle

\begin{abstract} 
Dissipation of magnetic fields in Hall plasma of neutron star crusts may power persistent high energy emission of a class of strongly magnetized  neutrons stars, magnetars.
We consider 
development of  a dissipative  tearing mode in  Hall plasma (electron MHD) and find that its  growth rate increases with the  wave number of perturbations, reaching a maximum value intermediate between 
resistive $\tau_r$ and Hall times $\tau_H$, $\Gamma\sim 1 / \sqrt{ \tau_r  \tau_H}$.
 We  argue that the  tearing mode may be the principal mechanism by which strong magnetic fields are  dissipated in magnetars on times scale of $\sim 10^4-10^5$ yrs powering the persistent X-ray emisison.
\end{abstract}

\section{Introduction}

Evolution of \NSs' magnetic fields is one of the principal issues in \NS research. On the one hand, observationally, there is no evidence that fields decay on long time scales, of the order of million years \cite{Hartman}. Yet, this constrains the dipolar component of a  magnetic field, generated, presumably,
by currents flowing in the superconductive core. The fate of a magnetic field in the crust may be different. It is tempting to relate an   X-ray activity of magnetars, strongly magnetized \NSs, to the decay of  a crustal magnetic field
created by a dynamo action during the \NS birth \cite{TD93,TD95}.

One of the  central problems is how fast a magnetic field can be dissipated. 
Conventionally,  the magnetic field decay time scale is assumed to be of the order of the resistive time scale, given by the size of the system $L$ (of the current-carrying layer, to be more precise) and resistivity $\eta$, $\tau_r \sim L^2/\eta$. In fact,  plasma has a number of ways to dissipate magnetic field on time scales much shorter than  $\tau_r$
(otherwise we would never see a Solar flare, since the  resistive time scale in the Solar  corona is longer than the age of the Universe). Typically, fast magnetic field  dissipation   is achieved through formation of small scale structures, with correspondingly short dissipation time scales. One possibility to create small scale structures is through non-linear interaction of waves, either locally (in phase space) through  formation of a Hall cascade \cite{RG}, or non-locally due to large scale motions leading to "wave overturn" 
\cite{Kingsep,Shalybkov,Bulanov,Vainshtein,PR06}. 

Here we  discuss another way to form small scale dissipative current layers via development of a  tearing mode in Hall plasma. In Hall plasma (sometimes called electron MHD, EMHD below \cite{Kingsep}) ions are assumed  to be motionless, providing a neutralizing background for electron fluid. 
Tearing mode \cite{furth} is the principal resistive instability of current-currying plasma and is one of the key factors leading to  an explosive release of magnetic energy in Solar flares \cite{Priest81} and magneto-tail \cite{Galeev78}.  It is also  important for TOKAMAK discharges like sawtooth oscillations and  
major disruptions  \cite{Kadomtsev75}. Tearing mode in regime is an important ingredient  of  modern reconnection models \cite{Bulanov,Biskamp}. During the  development of the  tearing mode, 
perturbations  of a plasma equilibrium   lead to formation of 
current sheets where both
the time scales for diffusion may be short and, in addition, resistivity may be 
enhanced due to the  development of plasma turbulence (anomalous resistivity).  A  current  
layer  tends to be locally  unstable   to transverse, $\k \cdot \B_0=0$, perturbations (${\bf k} $ is the wave vector of perturbations and $\B_0$ is initial magnetic field). 
Qualitatively, divergence at points  $\k \cdot \B_0=0$ is related to the fact that \Alfven velocity becomes zero at these points, so that perturbations are effectively piling up. 
Since whistler waves, normal modes in Hall plasma, also have phase speed equal to zero at points where  $\k \cdot \B_0=0$, we might expect a somewhat similar behavior. This is indeed what we show below.

\section{Tearing mode in Hall plasma} 
\subsection{Tearing due to electron resistivity}

In deriving  growth rates for the  tearing mode in Hall plasma we follow a standard scheme of electron-ion plasma
\cite{woods}. On scales much larger that electron skin depth,
the equation of resistive  EMHD   is (for example \cite{Kingsep})
\be
\partial _ t \B = \eta \Delta \B - {c \over 4 \pi  e} \nabla \times    \left({\nabla  \times \B  \over n} \times \B \right)
\label{0}
\ee
where $\eta$ is  resistivity, $n$ is plasma density and other notations are standard.
In what follows we assume density to  be constant. 

We assume that initial magnetic field configuration is stable both on \Alfven time scale (implying a dynamical stability) 
and on Hall time scale. In doing so we wish to single out  the effects of the tearing mode as opposed to dynamical instabilities, Hall cascade etc.

 Let the initial magnetic field correspond to   current layer $\B=B_0 (f_x (z) {\bf e}_x +f_y(z) {\bf e}_y)$ with
$f_x \rightarrow \pm 1$  at  $z  \rightarrow \pm \infty$ and $f_x =0$ in the center of the layer $z=0$. 
(If the layer is force-free, then   $f_x^2 + f_y^2 =1$; this is {\bf not} required for   tearing  instability).
Next we impose perturbation of the field in the form
\be
\delta \B = \nabla \times {\bf \xi} 
\ee
where $\xi \propto e^{-i (\om t - k_x x -k_y y)}$ is vector potential. 
We foresee instability at points ${\bf k} \cdot \B_0=0$, so to simplify  mathematics, we assume $k_y=0$, so that 
a current  sheet  will be created at $z=0$. The equations for evolution of perturbations are
\ba &&
\om \xi_y ' + {c ^2 \k_x \om_B \over \om_p^2} \left( i k_x(    f_x' \xi_z + f_x  \xi_z')- \xi_y f_y^{\prime \prime} - \xi_x^{\prime \prime} f_x  -f_x ' \xi_x' -f_y' \xi_y' \right) 
+  
\nn &&
\eta\left(i( k_x^2 \xi_y' - \xi_y ^{(3)} ) - { c m \om_B \over e}  f_x ^{\prime \prime} \right) 
=0
\nn &&
\om (i k_x \xi_z - \xi_x')+ {c ^2 \k_x \om_B \over \om_p^2} \left( k_x^2 f_x \xi_y +\xi_y f_x ^{\prime \prime} - f_x
\xi_y ^{\prime \prime} \right) 
\nn &&
-\eta\left(   k_x ( k_x^2 + \xi_z^{\prime \prime}) + { c m \om_B \over e} f_y ^{\prime \prime} + i( k_x^2 \xi_x'  - \xi_x^{(3)})\right)
=0
\nn &&
-i k_x \om \xi_y+ {c ^2 \k_x^2 \om_B \over \om_p^2} \left( k_x f_x \xi_z + i \xi_y f_y' + i f_x \xi_x'\right)
+\eta k_x \left( k_x^2 \xi_y  - \xi_y ^{\prime \prime} \right)
=0
\label{main}
\ea
The two scales of interest  that appear in the equations are 
\ba &&
\delta = \sqrt{ \eta \over | \om| }
\nn &&
\delta ^\ast = \sqrt{ L | \om | \delta^2 \om_p^2 \over c^2 k_x \om_B}=
( L k_x \delta^2) { | \om | \over \om_w}
\label{delta}
\ea
where $\om_w =  (c k_x)^2 \om_B /\om_p^2$ is the frequency of whistler waves. 
Generally,  dispersion relation of whistler waves
$\om_w = { c^2 k k_x \om_B  / \om_p^2}= { c^2 k (\k \cdot {\bf \om_B }) / \om_p^2} $, where $k=\sqrt{k_x^2 +k_z^2}$. In an  inhomogeneous plasma of our choice
$\om_w =0$  at a point $z=0$.
For comparison, in e-i plasma $\delta ^\ast = \sqrt{ L \delta |\om|/\om_A}$, with $\om _A = k_x v_A$.
In the above equations $| \om | $ is an absolution value of the (generally complex) frequency $\om$. As it turns out,
real part of $\om$ is not important (see below), so $ | \om| \sim \Gamma$, where $\Gamma$ is a growth rate of instability.

A standard method in describing the evolution of tearing mode
is similar to the boundary layer problem.  It 
involves  separation of a current layer into  "bulk", where derivatives
are small and resistivity is not important,  and a   narrow "boundary layer"
where derivatives and resistivity may be large. 
Two different approximations are done in each layer -
 ideal and weakly varying plasma in the bulk
and a narrow resistive  sublayer, with two solutions matched continuously.

In the outside region of smooth field evolution, where we can neglect resistivity, the system (\ref{main})  reduces to 
\be
\xi_y ^{\prime \prime} - \left(k_x^2 -  { \om^2 \om_p^4 \over c^4 \om_B^2 k_x^2 f_x^2} - 
{\om \om_p ^2 f_y ' \over c^2 k_x \om_B f_x^2} +  {f_x ^{\prime \prime}  \over f_x} \right) \xi_y
\ee
We are interested in the slow evolution of plasma and thus neglect terms proportional to $\om$. The structure  of magnetic field in the ideal region is then described by 
\be
\xi_y ^{\prime \prime} =\left(k_x^2  +  {f_x ^{\prime \prime}  \over f_x} \right) \xi_y
\label{1}
\ee
This clearly shows that instability is driven  by $f_x ^{\prime \prime}$ term, the second derivative of magnetic field.

The simplest choice of the background field in (\ref{1}) is $f_x = \sin z/L$, $z< L$, in which case 
\be
\xi_y ^{\prime \prime} =\left(k_x^2  - {1 \over L^2}  \right) \xi_y
\label{2}
\ee
with solution
\be
\xi _y = C_1 \cos \left({ \sqrt{1-k_x^2 L^2} z\over L} \right)
+ C_2 \sin  \left({ \sqrt{1-k_x^2 L^2} z\over L} \right)
\ee
Outside of the current sheet, $z>L$,  $\xi_y = e^{-k_x z}$ and the matching gives
\ba &&
C_1 = e^{-k_x L} \left (\cos  \sqrt{1-k_x^2 L^2} + {k_x L \sin \sqrt{1-k_x^2 L^2}\over  \sqrt{1-k_x^2 L^2}}\right)
\nn &&
C_2 = e^{-k_x L} \left (- \sin \sqrt{1-k_x^2 L^2}+{k_x L \sin \sqrt{1-k_x^2 L^2}\over  \sqrt{1-k_x^2 L^2}} \right)
\ea
Thus, magnetic field $B_z = k_x \xi_y$ is continuous at $z=0$, but its derivative experiences a jump
\be
\Delta = \left[ \ln B_z \right] \sim {1  \over L}
\ee
For instability it is required that $\Delta > 0$. 
(Recall, for comparison,  that in electron-ion plasma $\Delta \sim 1/(k L^2)$).

In the resistive sub-layer, at $z \ll L$, we have $f_x \rightarrow 0\, , f_y' \rightarrow 0$, so that the $z$ component of (\ref{main}) gives
\be
\xi_y^{\prime \prime} = { 1+k_x^2 \delta ^2 \over \delta ^2}\xi_y
\label{sub}
\ee
where $\delta ^2 = \eta/\Gamma$, $\Gamma = {\rm Im} (\om)$. In the limit $k_x \delta \ll 1$ this gives
\be
\xi_y = \cosh z/\delta
\ee
where a boundary condition $\xi_y' =0 $ at $z=0$ was used. (Structure of the resistive sub-layer is the same as in electron-ion  plasma).
The inner and outer solutions
match at 
\be
\delta^\ast = \Delta \delta^2
\label{3}
\ee
where $\delta^\ast $ is the thickness of the resistive sub-layer.


From Eqns (\ref{delta}), (\ref{3})
we get the growth rate
\be
\Gamma= {c^2 k_x \delta^2 \Delta^2 \om_B \over
L \om_p^2}  = \left( {c ^2 k_x \eta \om_B \over L^3 \om_p^2} \right)^{1/2}
\label{Gamma}
\ee
This is consistent with Eq. 50 in Ref.  \cite{Bulanov}, and Eq. (68) in Ref. \cite{Fruchtman}; see also \cite{Gordeev,Hassam}.
If we  define phase velocity of whistler modes as  $v_w = k_x c^2 \om_B/\om_p^2$, then Eq. (\ref{Gamma})
takes the form 
\be
\Gamma= {1 \over \sqrt{ \tau_r \tau_w}}
\ee
where 
 resistive time scale is $\tau_r =L^2/\eta$ and whistler time scale $\tau_w = L/v_w$. 
 
Growth rate (\ref{Gamma}) increases as  $k_x^{1/2}$. This is in stark contrast to electron-ion  plasma, where growth rate decreases with $k_x$ (as $\propto k^{-2/5}$,  \eg \cite{woods} Eq.  7.123).
Thus, in Hall plasma the maximum growth rate is reached at $k_x \sim 1/L$, and becomes
\be
\Gamma = { c \sqrt{\eta \om_B} \over L^2 \om_p} = {1 \over \sqrt{ \tau_r \tau_H}}
\label{Gamma1}
\ee
where we defined Hall time
\be
\tau_H = { L^2 \om_p^2 \over c^2  \om_B} 
\ee
Relation (\ref{Gamma}) gives the growth rate of tearing mode in Hall plasma. Thus, the time scale for development of a tearing mode  is intermediate between the resistive and Hall time scales, and thus is much  shorter than resistive  time scale. This expression reminds of the maximum growth rate of tearing mode in electron-ion plasma,
which is intermediate between resistive and \Alfven time scales.

Growth rate (\ref{Gamma1}) may also  be written as
\be
\Gamma = \left( {c  \over L \om_p} \right)^2 {\om_B \over \sqrt{ S} }
\ee
where we introduced an effective Lundquist number associated with Hall time 
$ S= v_H L/\eta$, $v_H=L/\tau_H$

\subsection{Tearing due to  inertial resistivity}

In  a highly conducting plasma, electron inertia may play a role of resistivity, providing a relation between the electric field and the current. It is  this regime that most works on tearing mode in Hall plasma addressed so far \cite{Bulanov}. Since in this case the width of the current sheet may become comparable to electron skin depth, one has to take into account additional terms in the Ohm's law and it's consequence, Eq. (\ref{0})  If we neglect this possibility (this is justified since we get correct expression for the growth rate, see below), the effective resistivity in the collisionless regime becomes 
$
\eta_{eff} = \Gamma  \left( {c  / \om_p} \right)^2 
$ \cite{Shivamoggi,Bulanov}. Thus, inertial resistivity dominates when $ \Gamma (c/\om_p)^2 > \eta$, which can be written as
$(c / L \om_p)^2 \sqrt{\tau_r/\tau_w}$.
 
The  corresponding tearing mode growth rate  is
\be
\Gamma = (k_x L) \left( {c \over L \om_p}\right)^4 \om_B = \Delta^{\prime 2} { c^4 k_x \om_B \over L \om_p^4}
\ee
This is the same  as the  growth for electron tearing mode cited in Ref.  \cite{Avinash}.
The maximum rate is reached at $k \sim 1/L$
\be
\Gamma_{max}= \left( {c \over L \om_p}\right)^4 \om_B 
\label{GammaIn}
\ee
The ratio of resistive (\ref{Gamma1}) and inertial  (\ref{GammaIn}) maximum  growth rates is
$ (c/L\om_p)^2 \sqrt{S}$. 

\section{Application to neutron stars}

High energy emission of strongly magnetized neutron stars (magnetars) is powered by dissipation of magnetic field  \cite{TD95,woods06, Lai06}. Strong magnetic fields, of the order of $10^{14}$ G, 
may be created  by a 
dynamo mechanism, \eg of the $\alpha-\Omega$ type,  operating at birth of neutron stars  \cite{TD93}. 
After $\sim $ 100 secs from  the birth,  the crust of the \NS solidifies. This time is much longer than \Alfven crossing time $\sim 0.1 -1 $ s, so that after the end of the turbulent motion and before the crust formation, magnetic field in a star should evolve to some minimum energy state allowed by the system.

After the crust solidifies, the evolution of magnetic field will proceed due to non-dissipative Hall effect, Ohmic resistivity, and, as we argue in this paper, due to  development of tearing instability.
In order to estimate the  corresponding   times scales, we use resistivity $\eta = c^2/ 4 \pi \sigma$ with conductivity
$
\sigma \sim 10^{25} \rho_{12}^{2/3}
$
\cite{Cummings} evaluated at the neutron drip point $\rho \sim 10^{12} $ g/cm$^{-3}$.
The  Hall $\tau_H$, resistive $\tau_r$ and tearing time scales (given by inverse of the tearing mode growth rate (\ref{Gamma})) become
\ba &&
\tau_H =4 \times 10^3 L_4^2  B_{14} ^{-1} 
\rho_{12} \, {\rm yrs}
\nn &&
\tau_r = 4 \times 10^5 L_4^2  \rho_{12}^{2/3} \, {\rm yrs}
\nn &&
1/\Gamma = 4 \times 10^4  L_4^2 \rho_{12}^{5/6} B_{14}^{-1/2}  \, {\rm yrs}
\label{time}
\ea
where a standard subscript notation, \eg $L_4 =(L/10^4{\rm cm} )$, has been adopted. 
This indicates that growth rate of the tearing mode in Hall plasma is of the same order as the activity time of magnetars.
Also, for higher magnetic field the  instability time is shorter while corresponding magnetic energy is larger, consistent with the fact that only high field \NSs exhibit magnetar-like activity. One can also check that in neutron star crusts resistive effects well dominate over inertial ones.

We envision a magnetar  scenario similar to the ones proposed in Refs \cite{tlk,bt}. After the turbulence seizes, initially stable magnetic field configuration forms which is  then  dissipated due to development of tearing mode  on time scale $\sim 10^4-10^5$  yrs. This leads to Lorentz force disbalance in the crust,  which initially can be   compensated by crust tensile strength, but  eventually  is    released through crust deformation  leading  to bursts  and  flares. (Release of crustal stress can be either  through cracks \cite{TD93}  or through plastic deformations \cite{lyut06}).

Since small scale currents are dissipated on shorter time scales, while larger currents take longer to dissipate, we expect that {\it giant flares are more common in  older magnetars}. 
 (This is also a natural consequence of the torus formation  \cite{Braithwaite}, in which case small scale magnetic fields are dissipated soon after the birth of a \NS).
 This may provide a resolution to the energy budget problem for magnetar giant flares: the giant flare of SGR 1806 - 20 emitted $2 \times 10^{46}$ ergs in high energy $\gamma$-rays \cite{palmer}, and associated mechanical energy is expected to be even larger. If a typical giant flare recurrence time is $\sim 30$ yrs (the period of active monitoring of high energy sky) and activity time is several thousand years (characteristic age) the magnetic field, which powers the flares, has to be {\it larger} than $3 \times 10^{15}$ G, which is on the verge of being uncomfortably large.

\section{Conclusion}

In this paper we first considered development of a resistive  tearing mode in Hall plasma. 
 Tearing mode develops on time scale intermediate  between Hall and resistive time scales. Qualitatively, this might have been expected in analogy with the tearing mode in electro-ion plasma. 
Unlike the case of electron-ion plasma, the growth rate (\ref{Gamma}) increases with $k$, reaching  maximum for $k\sim 1/L$, the thickness of the current layer. This is due, qualitatively, to the dispersion of whistler modes, for which phase velocity increases with $k$. 

 We then argue that the  development of the tearing mode in the crust of strongly magnetized neutron stars, magnetars, powers their high energy emission on time scales $\sim 10^4 -10^5$ years. One of the principal uncertainty is 
the length scale  $L$ of the current sheets created by the dynamo, as all the time scales are strongly dependent on it,
Eq. (\ref{time}). For our choice of parameters  the Hall time scale is $ \tau_H \sim 10^4$ yrs. According to Ref. \cite{Cummings}, this, in fact,  applies to a very broad density range, $\rho \sim 10^{11} - 10^{13}$ g cm$^{-3}$.  Ohmic decay times are strongly dependent on the temperature, with the value that we used, $\tau _R \sim 10^6$ yrs., giving a reasonable approximation.


Overall, the
evolution of a  crustal magnetic field is bound to be a combination of a  dissipative tearing mode and  a nondissipative Hall cascade. 
 The Hall turbulent cascade can create small scale structures which in turn become dissipative \cite{RG}. The time it takes for a cascade to propagate down to dissipative scale is, typically the large eddy overturn time times the logarithm of the ratio of outer to inner scales,
 $\sim \tau_H \ln L/L_{in} \gg \tau_H $ \cite{Zakharov} , where $L_{in}$ is the dissipative (resistive) scale. As the ratio
 $ L/L_{in}$ is large, it takes many Hall times to dissipate magnetic energy. Numerical simulations \cite{HollerbachRudiger}
 indeed seem to indicate that  the transfer of energy to the higher harmonics is not sufficient to accelerate significantly the decay of the original field. 
 
Relative importance of Hall cascade and tearing mode depend on details of 
the structure of magnetic field before solidification of the crust.
This   is  yet an unresolved  issue, as is exemplified by a long standing problem of  stability of magnetic field in stars  \cite{Prendergast,FlowersRuderman,Braithwaite,Broderick,Reisenegger07}. 
It appears that evolution at intermediate times (longer than \Alfven crossing time but shorter than   dissipative times) depends on magnetic  helicity, either at the end of the turbulent phase or due to helicity imbalance due to loss through the surface). For certain parameters, internal magnetic field relaxes to a complicated torus-like form with comparable toroidal and  poloidal magnetic fields.



Our results provides an elegant explanation to observation in Ref. 
 \cite{RheinhardtGeppert02}  (see also \cite{Pons}) where it is found that formation of  dissipative  current sheets 
requires  at least quadratic dependence of the  background field on the position.  Eq. (\ref{1}) clearly shows that instability is driven
by the second spacial  derivative of the field. Thus, what is calculated in Refs  \cite{RheinhardtGeppert02} is  a tearing mode. 
Finally, in Ref.
\cite{Vainshtein}  a somewhat related problem was  considered, the  formation of a current sheet  in  presence of a steep density gradient. In contrast, the tearing mode considered in this paper does not require  density gradient, all that is needed is an inhomogeneous  magnetic field. In addition, limitation of purely toroidal field, assumed in \cite{Vainshtein} seems to overestimate the dissipation
\cite{HollerbachRudiger04}.

Finally, the model presented here may be directly probed in the laboratory, in particular at the field-reversed configuration experiment at UCLA \cite{stenzel}. Tearing and formation of magnetic islands, a typical consequence of the nonlinear  development of  the tearing mode, are clearly seen
(\eg Fig 9 in Ref  \cite{stenzel}).

I would like to thank Andreas Reisenegger for his most valuable comments and numerous discussions. I also would like to thank  James Drake, Rainer Hollerbach, Ulrich Geppert, Matthias Reinhardt, Gennady Shvets and  Dmitri  Uzdensky
\bibliographystyle{prsty}
\bibliography{HallNS}

\end{document}